\documentclass[a4paper,12pt]{article}
\usepackage[T2A,OT1]{fontenc}
\usepackage[cp866]{inputenc}
\usepackage[dvips]{graphicx}

\title{\bf Unitarity of the tree approximation to
the Glauber AA amplitude
for  large A}
\author{M.A.Braun and A.V. Krylov\\
Dep. of High Energy physics,
 Saint-Petersburg State University}
\date{}
\def\beq{\begin{equation}}
\def\eeq{\end{equation}}

\begin{document}
\maketitle
\medskip

\fontencoding{OT1}\selectfont
\begin{abstract}
The nucleus-nucleus Glauber amplitude in the tree approximation
is studied for heavy participant nuclei. It is shown that, contrary to
previous published results, it is not unitary for realistic values
of nucleon-nucleon cross-sections.
\end{abstract}

\section{Introduction}
Scattering on the nuclei is commonly studied in the Glauber approximation,
which can be rigorously derived in Quantum Mechanics provided the
transverse momenta transferred to the projectile are much smaller than
its longitudinal momentum. With certain reservations it can be generalized
to the high energy region where the elementary nucleon-nucleon
(NN) amplitudes become predominantly inelastic. For the nucleon-nucleus
(NA) scattering the Glauber approximation has a transparent probabilistic
interpretation. If the target nucleus is heavy, with atomic number $A>>1$,
the Glauber formula acquires a simple eikonal form, which clearly shows
that the resulting amplitude is unitary, that is its modulus is smaller
than unity at fixed impact parameter.

\begin{figure}
\hspace*{1 cm}
\begin{center}
\includegraphics[width=12cm]{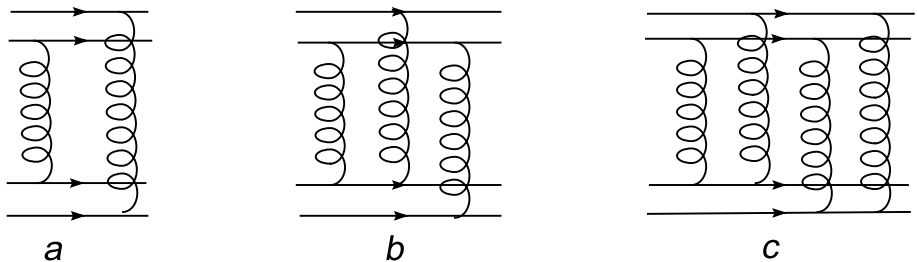}
\caption{Examples of disconnected ($a$), tree ($b$) and loop ($c$)
diagrams for the Glauber AB amplitude with $A=B=2$}
\end{center}

\smallskip
\label{fig}
\end{figure}

With the advent of collider experiments nucleus-nucleus (AB) scattering
becomes an important physical object. The Glauber approximation can be
easily generalized to the AB case and it was in fact done very long ago.
The Glauber formula for AB scattering looks very similar to the hA case.
At fixed impact parameter $b$ the scattering matrix $S$ is assumed to be a
product of  nucleon-nucleon scattering matrices $s$ averaged over the
transverse distributions of nucleons in both nuclei:
\begin{equation}
S(b)=\Big<\prod_{i=1}^A\prod_{k=1}^Bs(b-x_i+x'_k)\Big>_{A,B},
\label{sab}
\end{equation}
where $x_i$ and $x'_k$ are the transverse coordinates of the nucleons in
the projectile and target nuclei respectively and in absence of correlations
in both nuclei  averaging $<...>_{A,B}$ means
\begin{equation}
\Big<F(x_i,x'_k)\Big>_{A,B}=\int \prod_i^A\prod_k^Bd^2x_id^2x'_k
\Big(T_A(x_i)T_B(x'_k-b)F(x_1,...x_A, x'_1,...x'_B)\Big).
\label{average}
\end{equation}
Here $T_A(x)$ and $T_B(x')$ are the standard nuclear profile functions
normalized to unity.
However in contrast to the NA case the content of the Glauber formula
for AB scattering turns out to be much more complicated.
Presenting in the standard manner the NN scattering matrix
\[
s(b)=1+ia(b),
\]
where $a$ is the NN scattering amplitude,
one obtains from (\ref{sab}) a set of terms corresponding to different
ways the nucleons from the projectile and target may interact with
each other. Each of these terms may be illustrated by simple diagrams
indicating these interaction. Some examples are shown in Fig. 1. for
two pairs of interacting nucleons in the projectile and target.
One observes that in contrast to hA case the diagrams may contain
disconnected parts (Fig.\ref{fig}.$a$) and, most important,
loops (Fig.\ref{fig}.$c$), which involve
internal integrations over transferred transverse momenta and thus NN
amplitudes for non-zero transferred momenta.

Loop contributions depend
not only on the total NN cross-sections as the tree diagrams but also on
the form of the differential NN cross-section. Their calculation is
difficult and unreliable, so that in most applications to heavy nuclei
($A,B>>1$) loop contributions
are simply neglected. The typically used approximation is the so-called
optical approximation, which corresponds to taking into account only
the simplest contribution (a single NN interaction) for each connected
part. In the optical approximation
\begin{equation}
S^{opt}(b)=e^{F^{opt}(b)},
\label{sopt}
\end{equation}
where the optical eikonal $F^{opt}(b)$ is
\begin{equation}
F^{opt}(b)=iaAB\int d^2xT_A(x)T_B(b-x).
\label {eikopt}
\end{equation}
Its advantage is simplicity and obvious unitarity.
The natural question, which has been long discussed in literature,
is the quality of the approximation which neglects loops
(the tree approximation) for the case $A,B>>1$ and in
particular its relation to the optical approximation. In ~\cite{pak}
in the limit $A,B>>1$ a closed formula was obtained for the tree
approximation to the Glauber amplitude, different from the
optical approximation but also unitary
in the above mentioned sense, that is with a modulus smaller than unity
at fixed impact parameter.
Their result gives the tree eikonal in the form
\[
F^{~\cite{pak}}(b)=\frac{i}{a}\int d^2xf(\gamma_A,\gamma_B),
\]
where
\[
\gamma_A(x)=-iaAT_A(x),\ \ \gamma_B(x)=-iaBT_B(b-x),
\]
\[
f(\gamma_A,\gamma_B)=\sum_{l=1}^k(-1)^{l+1}(u_l+v_l+u_lv_l)
-\gamma_A-\gamma_B
\]
and $u_l$ and $v_l$ are the $k$ solutions of the transcendental equations
\begin{eqnarray*}
u=\gamma_Ae^{-v},\\ v=\gamma_Be^{-u}.
\end{eqnarray*}
The number of solutions may be $k=1$ or $k=3$. In the latter case it is
assumed that $u_1>u_2>u_3$ and $v_1<v_2<v_3$.
One can check by numerical studies that the resulting $F(b)$ is
always negative, so that $|S(b)|\leq 1$.
However later in ~\cite{kaidalov} it was
claimed that in fact at $A,B>>1$ the sum of all tree diagrams is exactly
given by the optical approximation.

The aim of the present study is to resolve this contradiction for the
asymptotic of the tree approximation to the Glauber AB amplitude for
large $A$ and $B$. Our results are different from both ~\cite{pak} and
~\cite{kaidalov}. Unfortunately they are also much gloomier.
Namely we find that at $A,B>>1$
the sum of all tree diagrams inevitably becomes non-unitary, so that
taking loops into account is absolutely necessary for the physically
meaningful amplitude. Thus the optimistic hope that the dependence of the
AB amplitude on the behaviour of the NN amplitude at non-zero
momentum transfers is hardly probable. ~\cite{pak} is unfortunately not
justified. The reason why our results are different from the previous ones
lies in the details in which simplifications related to the asymptotic at
$A,B>>1$ are made, in particular in the not completely accurate
use of the saddle point method in the previous derivations. As we shall see
apart from the saddle points which lead to the quasi-optical approximation
there exist others which may give the dominant contribution and destroy
unitarity.

Our study will be based on the equation which expresses in a compact manner
the sum of all tree diagrams, obtained by one of the authors some 30
years ago
~\cite{braun}. The good quality of this equation is that it is valid for
arbitrary finite (and even small) values of $A$ and $B$ and therefore
presents an adequate starting point to investigate the asymptotic at large
$A$ and $B$. To simplify we shall limit ourselves with the case
of collision of two identical nuclei
$A=B$. Since the problem we address does not depend on the form of
the transverse distribution $ T_A(x)$ we further simplify our study
by assuming that $T(x)$ does not depend on $x$ inside the nucleus:
\begin{equation}
T_A(x)=\theta(R_A-|x|)\frac{1}{\pi R_A^2}        
\label{profile}
\end{equation}
where $R_A=A^{1/3}R_0$ is the radius of the nucleus.

\section{The tree amplitude for AB scattering in the Glauber approximation}
\subsection {General $A$, $B$ and $b$}
In ~\cite{braun} an expression was derived for the sum of all tree
diagrams for the $S$ matrix in the Glauber approximation to nucleus-nucleus
scattering valid at arbitrary finite atomic numbers $A$ and $B$
of colliding nuclei. At  a given impact parameter $b$
\begin{equation}
S(b)=\frac{A!B!}{4\pi^2i^{A+B+2}}\oint\frac{d\tau d\tau'}
{\tau^{A+1}{\tau'}^{B+1}}e^{i(\tau+\tau')-Z(b,\tau,\tau')}.
\label{def_S_ot_b}
\end{equation}
Here
\begin{equation}
Z=-i\int(d^2x)
W\Big(i\tau T_A(x),i\tau' T_B(x')\Big)
\label{def_Z}
\end{equation}
and
\[
(d^2x)=d^2xd^2x'\delta^2(b-x+x').
\]
As indicated,
$a$ is the nucleon-nucleon forward scattering
amplitude and $T_A(x)$ and $T_B(x')$ are the nuclear profile functions at
transverse coordinates $x$ and $x'$ respectively. $W$ is the
effective classical action for the effective quantum theory of two fields
$u$ and $v$
\[
W(\rho,\rho')=Y\Big(u(\rho,\rho'),v(\rho,\rho')\rho,\rho', \Big),
\]
where
\[
Y=\frac{1}{a}vu-i\rho(e^{-u}-1)-i\rho'(e^{-v}-1)
\]
and $u$ and $v$ satisfy a pair of transcendental equations
\begin{equation}
u=-ia\rho'e^{-v},\ \ v=-ia\rho e^{-v}.
\label{eqphi}
\end{equation}
Here $\rho=i \tau T_A(x)$ and $\rho'=i\tau' T_B(x')$. 

A detailed derivation of this formula can be found in ~\cite{braun}.
For convenience it is briefly reproduced in Appendix.
In this section  for the simplified case
of constant $T_A$ and $T_B$ inside the nucleus we transform this formula
to the form suitable for our analysis at large $A$ and $B$.

It is trivial to see that $W$ is different from zero only in the overlap
region. Indeed for $T_A=0$ and so $\rho=0$ we have $v=0$  and for
$T_B=0$ and so $\rho'=0$ we have $u=0$. In both cases $Y=0$. Therefore
integration over $x$ and $x'$ is extended over the overlap
region only. For constant $T_{A,B}$ it gives precisely the area of the
overlap region $G(b)$. Taking this into account and expressing $W$ via $Y$
we find 
\begin{equation}
S(b)=\frac{A!B!}{4\pi^2i^{A+B+2}}\oint\frac{d\tau d\tau'}
{\tau^{A+1}{\tau'}^{B+1}}e^{i(\tau+\tau')}
e^{i\frac{G(b)}{a}(vu+\kappa\tau(e^{-u} -1)+
\kappa'\tau'(e^{-v} -1)},
\label{defs1}
\end{equation}
where
\begin{equation}
\kappa=\frac{a}{\pi R_A^2},\ \ \kappa'=\frac{a}{\pi R_B^2}
\label{defkappa}
\end{equation}
and equations (\ref{eqphi}) become
\begin{equation}
u=\kappa'\tau' e^{-v},\ \ v=\kappa\tau e^{-u}.
\label{equv}
\end{equation}
Separating in the exponent in (\ref{defs1}) the terms proportional to $\tau$
or $\tau'$ we rewrite Eq. (\ref{defs1}) as
\begin{equation}
S(b)=\frac{A!B!}{4\pi^2i^{A+B+2}}
\oint\frac{d\tau d\tau'}
{\tau^{A+1}{\tau'}^{B+1}}e^{i\tau \left(1-\frac{G(b)}{\pi R_A^2}\right)}
e^{i\tau'\left(1-\frac{G(b)}{\pi R_B^2}\right)}
e^{i\frac{1}{a}G(b)(uv+u+v)}.
\label{defs2}
\end{equation}

To avoid solving transcendenal equations (\ref{equv})
we pass in (\ref{defs2}) to the integration over $u$ and $v$,
since it is trivial to express $\tau$ and $\tau'$ via $u$ and $v$ from
(\ref{equv}) but not {\it vice versa}. To do this we have to find the Jacobian.
We denote $\xi=\kappa\tau$ and $\eta=\kappa'\tau'$
Direct differentiation of (\ref{equv})  gives
\[
\frac{\partial u}{\partial \xi}=-u\frac{\partial v}{\partial \xi},\ \
\frac{\partial u}{\partial \eta}=\frac{u}{\eta}-u\frac{\partial v}
{\partial \eta},\ \
\frac{\partial v}{\partial \xi}=\frac{v}{\xi}-v\frac{\partial u}
{\partial \xi},\ \
\frac{\partial v}{\partial \eta}=-v\frac{\partial u}{\partial \eta}.
\]
From these equation we immediately  obtain
\[
\frac{\partial u}{\partial \eta}=\frac{u}{\eta (1-uv)},\ \
\frac{\partial v}{\partial \xi} =\frac{v}{\xi (1-uv)},\ \
\frac{\partial u}{\partial \xi}=-\frac{uv}{\xi (1-uv)},\ \
\frac{\partial v}{\partial \eta} =-\frac{uv}{\eta (1-uv)}.
\]
As a result we find the Jacobian
\[
J=\frac{\partial(u,v)}{\partial(\xi,\eta)}=-\frac{uv}{\xi\eta (1-uv)}.
\]
At small $x$ and $y$ we obviously have $u\sim \eta$ and $v\sim \xi$.
So choosing the initial contours in $x$ and $y$ around the origin
small enough we find that integrations over $u$ and $v$ will go also
around small contours around the origin, which  can then be transformed
unless we come across some singularities in $u$ and $v$. Expressing $\tau$
and $\tau'$ in terms of $u$ and $v$ as
\[
\tau=\frac{1}{\kappa}ve^u,\ \ \tau'=\frac{1}{\kappa'}ue^v
\]
we transform Eq. (\ref{defs2}) into
\[
S(b)=\frac{A!B!}{4\pi^2i^{A+B+2}}\kappa^A{\kappa'}^B
\oint\frac{du dv}
{u^{A+1}v^{B+1}}(1-uv)e^{-Au-Bv}\]\begin{equation}
e^{i\frac{1}{\kappa}ve^u \left(1-\frac{G(b)}{\pi R_A^2}\right)}
e^{i\frac{1}{\kappa'}ue^v\left(1-\frac{G(b)}{\pi R_B^2}\right)}
e^{i\frac{1}{a}G(b)(uv+u+v)}.
\label{eqdb}
\end{equation}
This formula is the starting point for our investigation.

\subsection{Case $A=B$ and $b=0$}
Our formula for $S(b)$ greatly simplifies in the case of central
collisions of identical nuclei, when $A=B$ and $b=0$.
In this case the complicated exponents in the first two exponentials in
(\ref{eqdb}) are absent and we find a simple expression
\begin{equation}
S(0)=\frac{(A!)^2}{4\pi^2i^{2A+2}}\kappa^{2A}
\oint\frac{du dv}
{(uv)^{A+1}}(1-uv)e^{-A(u+v)}
e^{-\frac{1}{i\kappa}(uv+u+v)}.
\label{sb0}
\end{equation}
It is straightforward to find this $S$ matrix in the form of a
finite sum of terms. Integrations over $u$ and $v$
obviously give the coefficient before term $(uv)^A$ in the expansion of
the rest part of the integrand in powers of $u$ and $v$ (with factor
$(2\pi i)^2$ which cancels with the analogous factor in front of the
whole expression in (\ref{sb0})). So our problem reduces to the expansion
in powers of $u$ and $v$ of the three exponentials in (\ref{sb0}).
We find
\[
e^{\left(-\frac{1}{i\kappa}-A\right)u}
e^{\left(-\frac{1}{i\kappa}-A\right)v}
e^{-\frac{1}{i\kappa}uv}\]\[ =
\sum_{n_1,n_2,n_3}\frac{1}{n_1!n_2!n_3!}\left(-\frac{1}{i\kappa}\right)^{n_3}
\left(-\frac{1}{i\kappa}-A\right)^{n_1+n_2}u^{n_1+n_3}v^{n_2+n_3}=\]
\begin{equation}
=\sum_{n=0}\frac{1}{n!}\left(-\frac{1}{i\kappa}\right)^{n}
\sum_{n_1,n_2\geq n}\frac{u^{n_1}v^{n_2}}{(n_1-n)!(n_2-n)!}
\left(-\frac{1}{i\kappa}-A\right)^{n_1+n_2-2n}.
\end{equation}
Integration over $u$ and $v$ gives the term with $n_1=n_2=A$. Without the
factor $(2\pi i)^2$ and the one in front of the whole expression (\ref{sb0}) it is
\[
\sum_{n=0}^A\frac{1}{n![(A-n)!]^2}\left(-\frac{1}{i\kappa}\right)^{n}
\left(-\frac{1}{i\kappa}-A\right)^{2(A-n)}\]\begin{equation}=
\sum_{n=0}^A\frac{1}{n![(A-n)!]^2}\left(-\frac{1}{i\kappa}\right)^{2A-n}
(1-\gamma)^{2(A-n)}.
\label{sum1}
\end{equation}
where we defined
\begin{equation}
\gamma=-iA\kappa.
\label{defgamma}
\end{equation}
From this expression one has to subtract the second one which comes from
the term $-uv$ in the Jacobian. Obviously it gives the $(A-1)$th term in the
expansion of the integrand in powers of $u$ and $v$ and is obtained from
(\ref{sum1}) by the substitution $A\to A-1$.

Collecting all the factors we get
\[
S(0)=
(A!)^2 (1-\gamma)^{2A}\Big\{
\sum_{n=0}^A\frac{1}{n![(A-n)!]^2}\left(\frac{\gamma}
{A(1-\gamma)^2}\right)^n-\]\begin{equation}
\frac{\gamma^2}{A^2(1-\gamma)^2}
\sum_{n=0}^{A-1}\frac{1}{n![(A-1-n)!]^2}
\left(\frac{\gamma}{A(1-\gamma)^2}\right)^n
\Big\},
\label{sumfors}
\end{equation}
Note that in fact the
scattering amplitude $a$ is pure imaginary at high energies:
\[
a=\frac{i}{2}\sigma,
\]
where $\sigma$ is the total cross-section for pp collisions.
Threrefore $\gamma$ is positive and so each term in the two sums in
(\ref{sumfors}) is positive.
This expression gives a simple closed form for the $S$ matrix
for collisions of identical nuclei at $b=0$. The term with $n=0$ in the first
sum is independent of the scattering amplitude $a$ and equal to unity.
The rest terms give the amplitude with factor $i$.

Note that (\ref{sumfors}) contains powers  $(1-\gamma)^n=(1+Ai\kappa)^n$.
This is the origin of  difficulties related to AB scattering.
A similar formula for NA scattering contains powers  $(1+i\kappa)^n$.
Since $\kappa$ is small, of the order $A^{-2/3}$, summation over $n$ does
not violate unitarity. In contrast, for the nucleus case $\kappa$ is
substituted by $A\kappa$, which grows with $A$ as $A^{1/3}$. As a result
factors $(1-\gamma)^n$ grow like $A^{n/3}$ at high $A$ and $n$ and as
we shall discover make $|S|$ also grow. Unitarity remains valid only for
values of $\gamma$ just slightly above unity, which is certainly not
satisfied at high enough (and physically interesting)  $A$.

Eq. (\ref{sumfors}) makes it feasible to perform  numerical calculations
of the tree Glauber amplitude, since each of the two sums contains only
positive terms.
Numerical results demonstrate that $S(0)$ is unitary, that is
$|S(0)|<1$, at any value of $A$ only provided $0<\gamma<1.42$.
The last condition means that
\begin{equation}
\frac{A\sigma}{2\pi R_A^2}<1.42.
\label{cond}
\end{equation}
To compare, a similar condition for $hA$ scattering
\[
\frac{\sigma}{2\pi R_A^2}<1
\]
is always satisfied for large $A$. However in our case, with $R_A=R_0 A^{1/3}$
(\ref{cond}) reduces to
\begin{equation}
\frac{A^{1/3}\sigma}{2\pi R_0^2}<1.42.
\label{cond1}
\end{equation}
With $\sigma\sim 2\pi R_0^2$ it is always violated at physically relevant
large $A$.

\section{The asymptotic of the amplitude at $b=0$ and $A=B\to\infty$}
In this section we shall derive analytic asymptotic formulas which
agree with our numerical results.

We rewrite (\ref{sb0}) as
\begin{equation}
S(0)=\frac{(A!)^2}{4\pi^2i^2}\frac{\gamma^{2A}}{A^{2A}}
\oint du dv\Big(\frac{1}{uv}-1\Big)e^{AP(u,v)},
\label{newd0}
\end{equation}
where
\[
P(u,v)=-\ln(uv)-u-v+\frac{1}{\gamma}(uv+u+v).
\]
We estimate the asymptotic by the saddle point method.
The saddle point is determined by the equations
\begin{eqnarray}
P_u=-\frac{1}{u}-1+\frac{1}{\gamma}(v+1)=0,\nonumber\\
P_v=-\frac{1}{v}-1+\frac{1}{\gamma}(u+1)=0.
\label{deriv1}
\end{eqnarray}
The second derivatives are
\[
P_{uu}=\frac{1}{u^2},\ \ P_{vv}=\frac{1}{v^2},\ \ P_{uv}=\frac{1}{\gamma}.
\]
Eqs. (\ref{deriv1}) have two symmetric solutions
\[
u_1=v_1=\gamma,\ \ u_2=v_2=-1.
\]

{\bf 1. Saddle points $u_s=v_s=\gamma$}

Passing to variables $u-\gamma=i\xi$ and $v-\gamma=i\eta$ we rewrite
the integral
in (\ref{newd0}) in the vicinity of the saddle point as
\[
S(0)=\frac{(A!)^2}{4\pi^2}\frac{\gamma^{2A}}{A^{2A}}e^{AP(\gamma,\gamma)}
\Big(\frac{1}{\gamma^2}-1\Big)
\int d\xi d\eta e^{-\frac{A}{2\gamma^2}(\xi^2+2\gamma\xi\eta+\eta^2)}.
\]
Subsequent actions depend on the sign of eigenvalues of the matrix
\[
M=\frac{A}{2\gamma^2}\left(\begin{array}{cc}
                    1&\gamma\\
                    \gamma&1\end{array}\right).
\]
The eigenvalues are
\[
\frac{A}{2\gamma^2}(1+\gamma)\ \ {\rm and}\ \ \frac{A}{2\gamma^2}
(1-\gamma).
\]
We have to consider two cases $\gamma<1$ and $\gamma>1$.

{\bf 1.1} Case $\gamma<1$

In this case we can safely extend the integration regions in both
$\xi$ and $\eta$ to the whole real axis to find
\begin{equation}
I=\int d\xi d\eta e^{-\frac{A}{2\gamma^2}(\xi^2+2\gamma\xi\eta+\eta^2)}=
\frac{\pi}{\sqrt{\det M}}=\frac{2\pi \gamma^2}{A\sqrt{1-\gamma^2}}.
\label{inti1}
\end{equation}
We further use
\begin{equation}
A!=A^Ae^{-A}\sqrt{2\pi A}
\label{Stirling}
\end{equation}
and
\[
AP(\gamma,\gamma)=-A\ln \gamma^2-A\gamma+2A
\]
to finally find
\begin{equation}
S(0)=\sqrt{1-\gamma^2}e^{-A\gamma}.
\label{asym0}
\end{equation}

{\bf 1.2} Case $\gamma>1$

In this case we have to rotate one of the variables $\tilde{\xi}$
or $\tilde{\eta}$ which diagonalize matrix $M$ which corresponds to
eigenvalue $1-\gamma$ by angle $\pm\pi/2$. Then we get instead of
(\ref{inti1})
\[
I=\pm i\frac{2\pi \gamma^2}{A\sqrt{\gamma^2-1}}
\]
and for $S(0)$
\begin{equation}
S(0)=\mp i\sqrt{\gamma^2-1}e^{-A\gamma}
\end{equation}

{\bf 2. Saddle points $u_s=v_s=-1$}

In this case the prefactor $(uv)^{-1}-1$ vanishes at the saddle point
and we have to study it in the vicinity of the saddle point.
We put $u=-1-i\xi$ and $v=-1-i\eta$ to find
\[
\frac{1}{uv}-1=(1-i\xi-\xi^2)(1-i\eta-\eta^2)-1=-i\xi-i\eta-\xi^2-\eta^2
-\xi\eta.
\]
Since the leading term vanishes we have to expand the exponent up to
terms of the third order in $\xi$ and $\eta$
\[
P(u,v)=P(-1,-1)-\frac{1}{2}(\xi^2+\eta^2)-\frac{1}{\gamma}\xi\eta
+\frac{1}{3}i(\xi^3+\eta^3),
\]
so that in the vicinity of the saddle points the integrand in (\ref{newd0})
can be presented as
\[
\Big(\frac{1}{uv}-1\Big)e^{AP(u,v)}=-
e^{A(P(-1,-1)-\frac{A}{2}(\xi^2+2\xi\eta/\gamma+\eta^2)}Q(\xi,\eta),
\]
where the polynomial $Q(\xi,\eta)$ is
\[
Q(\xi,\eta)=\xi^2+\eta^2+\xi\eta-\frac{A}{3}
(\xi^4+\eta^4+\xi^3\eta+\xi\eta^3).
\]

So we find
\[
S(0)=-\frac{(A!)^2}{4\pi^2}\frac{\gamma^{2A}}{A^{2A}}e^{AP(-1,-1)}
\int d\xi d\eta
Q(\xi,\eta)e^{-\frac{A}{2}(\xi^2+2\xi\eta/\gamma+\eta^2)}
=-\frac{(A!)^2}{4\pi^2}\frac{\gamma^{2A}}{A^{2A}}e^{AP(-1,-1)} J,
\]
where
\[
J=j_0-\frac{A}{3}(j_1+j_2)
\]
with
\[
j_0= -\Big(\frac{\partial}{\partial\alpha}+
\frac{1}{2}\frac{\partial}{\partial\beta}\Big)J_0; \ \
j_1=\Big(\frac{\partial^2}{\partial\alpha^2}-
\frac{1}{2}\frac{\partial^2}{\partial\beta^2}\Big)J_0; \ \
j_2=\Big(\frac{1}{2}\frac{\partial^2}{\partial\alpha\partial\beta}\Big)J_0,
\]
\[
J_0=\int d\xi d\eta
e^{-\alpha(\xi^2+\eta^2)-2\beta\xi\eta}
\]
and
\[
\alpha=\frac{A}{2},\ \ \beta=\frac{A}{2\gamma}.
\]
In the exponent the matrix in variables $\xi,\eta$ is now
\[
M=\left(\begin{array}{cc}
                    \alpha&\beta\\
                    \beta&\alpha\end{array}\right)
\]

Consider the case $\gamma>1$ Then $\alpha>\beta$ and both
eigenvalues of matrix $M$ are positive.
Then we immediately get
\[
J_0=\frac{\pi}{\sqrt{\alpha^2-\beta^2}}, \ \
j_0=
\frac{2\pi \gamma^2(2\gamma-1)}{A^2(\gamma^2-1)^{3/2}}, \ \
j_1=\frac{12\pi\gamma^5}{A^3(\gamma^2-1)^{5/2}}, \ \
j_2=-\frac{12\pi\gamma^4}{A^3(\gamma^2-1)^{5/2}}
\]
Using also
\[
AP(-1,-1)=2A-\frac{A}{\gamma}
\]
and the asymptotic  (\ref{Stirling}) we finally find for $\gamma>1$
\begin{equation}
S(0)=-\frac{ \gamma^2}{A(\gamma+1)^{5/2}(\gamma-1)^{1/2}}
e^{A(2\ln \gamma-\frac{1}{\gamma})}.
\end{equation}
So for $\gamma>1$ $S(0)$ is always negative and its
modulus exponentially grows
with $A$ unless $\gamma^2<\exp(1/\gamma)$, that is
$\gamma<\gamma_0=1.4215$ when it exponentially falls.

The asymptotic for $\gamma<1$  requires additional rotation in
variables which diagonalize matrix $M$. Up to its sign it can be found
just by the analytic continuation from the case $\gamma>1$.
So for $\gamma<1$
\begin{equation}
S(0)=\pm i\frac{ \gamma^2}{A(\gamma+1)^{5/2}(1-\gamma)^{1/2}}
e^{A(2\ln \gamma-\frac{1}{\gamma})}.
\end{equation}

One observes that  for $\gamma<1$ the leading contribution comes
from the saddle point $u_s=v_s=\gamma$ and for $\gamma>1$ from the
saddle point $u_s=v_s=-1$, the latter contribution restricting the
region of $\gamma>1$ where unitarity is fulfilled

\section{Non-central collisions of identical nuclei}
We introduce $\gamma$ according to (\ref{defgamma}) and put
\begin{equation}
G(b)=\lambda(b)\pi R_A^2,\ \ 0\leq\lambda\leq 1
\label{deflambda}
\end{equation}
to rewrite Eq. (\ref{eqdb}) as
\begin{equation}
S(b)=\frac{(A!)^2\gamma^{2A}}{4\pi^2i^{2}A^{2A}}
\oint du dv\Big(\frac{1}{uv}-1\Big)e^{AP(u,v)},
\label{eqdb1}
\end{equation}
where now
\[
P(u,v)=-\ln (uv)-u-v+\frac{\lambda}{\gamma}(uv+u+v)+\frac{1-\lambda}{\gamma}
\Big(ve^u+ue^v).
\label{puv}
\]
The derivatives are
\[
P_u=-\frac{1}{u}-1+\frac{\lambda}{\gamma}(v+1)+
\frac{1-\lambda}{\gamma}\Big(ve^u+e^v\Big),
\]
\[
P_v=-\frac{1}{v}-1+\frac{\lambda}{\gamma}(u+1)+
\frac{1-\lambda}{\gamma}\Big(ue^v+e^u\Big),
\]
\[P_{uu}=\frac{1}{u^2}+\frac{1-\lambda}{\gamma}ve^u,\ \
P_{uv}=\frac{\lambda}{\gamma}+\frac{1-\lambda}{\gamma}\Big(e^u+e^v\Big).
P_{vv}=\frac{1}{v^2}+\frac{1-\lambda}{\gamma}ue^v,\ \
\]
As before we seek for symmetric stationary points $u_s=v_s$. Then we
obtain an equation
\[
(u_s+1)\Big(-\frac{1}{u_s}+
\frac{\lambda}{\gamma}+\frac{1-\lambda}{\gamma}e^{u_s}\Big)=0.
\]
We have the same solution $u_2=v_2=-1$ and a
new solution $u_1=v_1$ which satisfies
\[
-\frac{1}{u_1}+\frac{\lambda}{\gamma}+\frac{1-\lambda}{\gamma}e^{u_1}=0
\]
or
\begin{equation}
u_1=\frac{\gamma}{\lambda+(1-\lambda)e^{u_1}}.
\end{equation}
The actual value of $u_1$ for given $\lambda$ and $\gamma$ can only be found
numerically.

{\bf 1. Saddle points $u_s=v_s=u_1$.}

We find at the stationary point
\[
P=-2\ln u_1+2-2u_1+\frac{\lambda}{\gamma}u_1^2,
\]
\[
P_{uu}=P_{vv}=\frac{1}{u_1^2}+1-\frac{\lambda}{\gamma}u_1,\ \
P_{uv}=\frac{2}{u_1}-\frac{\lambda}{\gamma}.
\]
The determinant of the quadratic form in $u,v$ in the exponent turns out to be
\[
\det M=\frac{A^2}{4}(1-u_1^2)\Big[\frac{1}{u_1^4}-
\Big(\frac{1}{u_1}-\frac{\lambda}
{\gamma}\Big)^2\Big].
\]
It is positive for $u_1\leq 1$ and arbitrary $\lambda\leq 1$.
So for $u_1\leq 1$ we find the asymptotic
\begin{equation}
S(b)=\sqrt{\frac{1-u_1^2}{1-u_1^2(1- u_1\lambda/\gamma)^2}}
e^{-A\Big(2u_1-\frac{\lambda}{\gamma}u_1^2+2\ln\frac{u_1}{\gamma}\Big)}.
\label{asymb}
\end{equation}
If $b=0$ and $\lambda=1$ then $u_1=\gamma$ and this asymptotic passes into
(\ref{asym0}).
The bracket in the exponent in (\ref{asymb}) is always positive
and diminishes with $\lambda$, which implies that
the asymptotic gets less falling with the growth of $b$.
At $b=2R_A$ and $\lambda=0$, when the nuclei only touch each other,
the exponent vanishes, which corresponds to $S(b=2R_A)=1$ as it should be.

If $u_1>1$ then the asymptotic can be obtained by analytic continuation of
(\ref{asymb}). It remain to be falling with $A$. However in some regions of
$\lambda$ and $u_1$ it becomes pure imaginary.

{\bf 2. Saddle points $u_s=v_s=-1$.}

At this saddle point we find
\[
P=2-\frac{\lambda}{\gamma}-\frac{2}{e}\frac{1-\lambda}{\gamma}.
\]
Together with the prefactor in (\ref{eqdb1}) it gives an exponential factor
in the asymptotic
\begin{equation}
e^{A\Big(2\ln\gamma-\frac{\lambda}{\gamma}-
\frac{2}{e}\frac{1-\lambda}{\gamma}\Big)}.
\label{expfac}
\end{equation}
It infinitely grows at $\gamma>\gamma_0(b)$ where
\[
2\gamma_0\ln\gamma_0=\lambda+\frac{2}{e}(1-\lambda).
\]
The value of $\gamma_0(b)$ steadily but slowly grows with $\lambda$ (see Table 1 ).
These values restrict the region in which the amplitude
remains unitary at $b>0$.
\vspace{0.5 cm}
\begin{center}
Table 1. $\gamma_0$ as a function of overlap $\lambda(b)$
\vspace{0.5 cm}

\begin{tabular}{|c| |c| |c| |c| |c| |c| |c|}
\hline
$\lambda$: & $0$ & $0.2$ & $ 0.4$ & $0.6$ & $0.8$ & $1.0$\\
\hline
$\gamma_0$: & $1.3211$ & $1.3416$ & $1.3620$ & $1.3820$ & $1.4019$ & $1.4215$\\
\hline
\end{tabular}
\smallskip
\end{center}

Note that at $b=2R_A$ and thus $\lambda=0$ the exponential factor
(\ref{expfac}) grows with $A$ unless $\gamma< \gamma_0(0)=1.3211$,
in spite of the fact that  $S(b=2R_A)=1$. This seeming contradiction
is resolved due to  vanishing of the prefactor at exactly $\lambda=0$.
As soon as $\lambda$ gets slightly greater than zero, the $S$-matrix
becomes very large and negative for $\gamma>\gamma_0(b)$. For example
for $\lambda=0.001$ and $\gamma=2.0$ one finds $S(b)=-0.174\cdot 10^5$,
which illustrates that the asymptotic becomes discontinuous at $b=2R_A$
in the limit of large $A$.

\section{Wrong ways to study the asymptotic}
In this section we illustrate how unwarranted applications of the saddle
point method can easily lead to incorrect results for the asymptotic of
the tree approxumation to the Glauber AB amplitude, in particular to
the results found in refs. ~\cite{pak} and ~\cite{kaidalov}.

One may try to study the asymptotic directly from the representation
(\ref{defs2}) without passing to variables $u$ and $v$. For $A=B$
it can be rewritten as
\begin{equation}
S(b)=\frac{{A!}^2}{4\pi^2i^{2A+2}}
\oint\frac{d\tau d\tau'}
{\tau^{A+1}{\tau'}^{A+1}}e^{i (1-\lambda)(\tau+\tau')}
e^{i\frac{\lambda}{\kappa}(uv+u+v)},
\label{defs2a}
\end{equation}
where $u$ and $v$ are determined via $\tau$ and $\tau'$ by Eqs.
(\ref{equv}) and
$\kappa$ and $\lambda$  are defined by (\ref{defkappa}) and
(\ref{deflambda}) respectively.
The results of ~\cite{pak} are obtained if we separate a factor in
the integrand
\[
e^{i(\tau+\tau')-A\ln(\tau\tau')}
\]
and consider it as the only rapidly changing one  at $A\to \infty$.
Then the saddle points are
\[
i\tau=i\tau'=A.
\]
Putting these values in the integrand one obtains
the asymptotic of the $S$-matrix in the form
\[
S(b)=e^{F(b)},
\]
where
\[
F(b)=\frac{i\lambda}{\kappa}(uv+u+v)-2A\lambda
\]
and $u$ and $v$ are determined by
\[
u=-iA\kappa e^{-v},\ \ v=-iA\kappa e^{-u}.
\]
This is the  result of ~\cite{pak} for the case of $A=B$ and
constant profile functions.

However this derivation is too crude, since it neglects
the $A$ dependence of the action. A more elaborate derivation
based on variables $u$ and $v$ leads to a pure optical approximation.
Introducing for $A=B$
integration variables $t$ and $t'$ defined as
\[
i\tau=At,\ \  i\tau'=Bt'
\]
we find
\begin{equation}
S(b)=\Big(\frac{A!}{A^A}\Big)^2\frac{1}{4\pi^2}
\int\frac{dtdt'}{tt'}e^{AP(t,t')}
\label{dviat},
\end{equation}
where
\[
P_1=-\ln (tt')+(t+t')+\frac{\lambda}{\gamma}
(vu+u+v-r-r')
\]
with $r=\gamma t$ and $r'=\gamma t'$,
$u$ and $v$ determined via $t$ and $t'$ by equations
\begin{equation}
u=r' e^{-v},\ \ v=r e^{-u}
\label{detuv}
\end{equation}
and  $\gamma$ defined by Eq.
(\ref{defgamma}).
Applying the saddle point method to the integral (\ref{dviat})
one searches for  saddle points from equations
\[
\frac{\partial P_1}{\partial t}=\frac{\partial P_1}{\partial t'}=0.
\]

Elementary calculations using Eqs. (\ref{detuv}) give
\[
\frac{\partial P_1}{\partial t}=1-\frac{1}{t}-\lambda +\frac{\lambda}{\gamma}
\frac{v}{t},\ \
\frac{\partial P_1}{\partial t'}=1-\frac{1}{t'}-\lambda +\frac{\lambda}{\gamma}
\frac{u}{t'}
\]
If additionally $\lambda=1$ (central collisions) then the first and third
terms cancel and the
saddle points are found to be
\[
u=v=\gamma.
\]
They are the same as we had earlier, after passing to variables
$u$ and $v$. They lead to the optical approximation (see (\ref{asym0})).
However we do not find the other pair of saddle
points $u=v=-1$. The reason is that in variables
$t$ and $t'$ this saddle point is transformed into a singularity
in the $t,t'$-plane present in the solutions of Eqs. (\ref{detuv}).
This singularity takes the leading role in the asymptotics at $\gamma>1$
and leads to the growth of the $S$-matrix and violation of unitarity.

\section{Conclusions}
Using  the simplified form of the profile functions, constant inside the
colliding nuclei, we have found that the set of tree diagrams for the
AB amplitude in the Glauber approximation is not unitary for heavy
particpants and realistic values of the NN cross-section.
This fact has been found analytically, using the saddle point method
in the adequate manner, and fully confirmed by the straightforward
numerical calculations. Unitarity is found to be fulfilled only if
the NN cross-section $\sigma$ is small and diminishes with $A$ (see
Eq. (\ref{cond1})). Previous optimistic results ~\cite{pak,kaidalov}
are found to be incorrect due to inadequate application of the saddle
point method.

From the practical point of view our results mean that the
treatment of the AB scattering in the standard Glauber approximation
must inevitably include loop diagrams and so depend on the form of the
differential NN cross-section at non-zero angles. It remains to be studied
how the situation changes if AB scattering is not described by the Glauber
formula but is rather formed by the exchange of  self-interacting
pomerons (like in the Regge-Gribov model). In this case  loops can
also be formed, which are usually neglected because formally they
are subdominant in the parameter $A^{-1/3}$. However it is not clear
if the resulting tree amplitude is unitary when the limitation on
the number of interacting nucleons at finite $A$ is correctly imposed.
We leave this problem  for future studies.

\section{Acknowledgments}
This work has been supported by grants RNP 2.1.1/1575
and RFFI 09-02-01327-a.

\section {Appendix. Derivation of the integral representation for S-matrix.}
Consider the contribution to the AB scattering amplitude with a given number
$c$ of connected parts and a given number of participant nucleons
$n_i$ ($n'_i$)  from the projectile (target) in the $i$-th connected part
($i=1,...,c$).
The standard derivation leads to the expression
\begin{equation}
i{\cal A}_{c,n_i,n'_i}(b)=N\frac{A!B!}{c!(A-n)!(B-n')!}
(ia)^l\prod_{i=1}^c\int d^2x_iT_A^{n_i}(x_i)T_B^{n'_i}(x_i-b).
\label{ap1}
\end{equation}
Here $l$ is the total number of interactions:
$l=\sum_{i=1}^cl_i$ where $l_i$ is the number of interactions in the
$i$-th connected part. Likewise $n=\sum_{i=1}^cn_i$ and $n'=\sum_{i=1}^cn'_i$
are the total numbers of participants in the projectile and target.
In the tree approximation we study $l_i=n_i+n'_i-1$, so that the number of
interactions is uniquely determined by the number of participants.
The symmetry factor $N$ arises because with a given number of participant
nucleons in each connected part there may be several terms which give identical
contribution. To find this factor one may consider an auxiliary
zero-dimensional quantum field theory with a generating functional
\[
Z=\int D\phi D\phi^{\dagger} e^{iY(\phi,\phi^{\dagger},\rho,\rho')},
\]
where
\[
Y=a^{-1}\phi^{\dagger}\phi-i\rho\Big(e^\phi-1\Big)
-i\rho'\Big(e^{\phi^{\dagger}}-1\Big).
\]
This theory will generate the same diagrams as the Glauber expression
(\ref{ap1}) except that coordinate dependent densities
$T_A(x)$ and $T_B(x'-b)$ will be substituted by constants
$\rho$ and $\rho'$. The sum of all connected diagrams will be given by the
effective action $W$ obtained after integrating out the fields $\phi$ and
$\phi^{\dagger}$. It will depend on powers of $\rho$ and $\rho'$ corresponding
to numbers of fields entering in different connected parts
\[
iW(\rho,\rho')=\sum_iC_{n_i,n'_i}\rho^{n_i}{\rho'}^{n'_i}.
\]
The coefficients $C_{n_i,n'_i}$ are just the symmetry factors which
should be taken into account in Eq. (\ref{ap1}) except that they also
include the corresponding number of amplitudes $ia$. So we obtain
\[
i{\cal A}_{c,n_i,n'_i}(b)=\frac{A!B!}{c!(A-n)!(B-n')!}
\prod_{i=1}^cC_{n_i,n'_i}\int d^2x_iT_A^{n_i}(x_i)T_B^{n'_i}(x'_-b).
\]
Since we are interested only in tree diagrams, the effective action $W$
should be taken in the classical approximation:
\[
W(\rho,\rho')=Y(\phi(\rho,\rho'),\phi^{\dagger}(\rho,\rho'),\rho,\rho'),
\]
where the classical fields are determined from a pair of transcendental
equations
\[
\phi=ia\rho' e^{\phi^{\dagger}},\ \ \phi^{\dagger}=ia\rho e^\phi.
\]

Next we sum over all posible values of $n_i$, $n'_i$ and $c$.
To sum over $n_i$ and $n'_i$ we present
\[
C_{n,n'}=\oint\frac{dzdz'}{(2\pi i)^2z^{n+1}{z'}^{n'+1}}iW(z,z').
\]
Summations over $n_i$ and $n'_i$ in the integrand factorize in two
factors
\[
f_A(u_i)=\sum_{n_i}\frac{A!}{(A-n)!}
\prod_{i=1}^cu_i^{n_i},\ \ n=\sum_{1}^cn_i,
\]
where $u_i=T_A(x_i)/z_i$ and a similar factor for the target B. We present
each $u_i^{n_i}$ as
\[
u^n=\frac{1}{un!}\int_0^\infty dtt^ne^{-t/u}
\]
to find a sum in the integrand
\begin{equation}
\sum_{n_i}\frac{A!}{(A-n)!}\prod_{i=1}^c\frac{t_i^{n_i}}{n_i!}
=\Big(1+\sum_{i}^ct_i\Big)^A -\ \ {\rm zero\ \ terms}.
\label {apsum1}
\end{equation}
The subtraction terms eliminate contributions from the $i$-th connected part
when all $n_i=0$. The right-hand side of Eq. (\ref{apsum1}) can be
represented as a contour integral
\[
\frac{A!}{2\pi i^{A+1}}\oint \frac{d\tau}{\tau^{A+1}}e^{i\tau}
\prod_{i=1}^c\Big(e^{i\tau t_i}-1\Big).
\]
Integration over all $t_i$ gives
\[
f_A(u_i)=\frac{A!}{2\pi i^{A+1}}\oint\frac{d\tau}{\tau^{A+1}}
\prod_{i=1}^c\Big(\frac{1}{1-i\tau u_i}-1\Big).
\]
The final expression for the amplitude is obtained after integrations over
all $z_i$ and summation over the number of connected parts $c$, which are
realized in a straightforward manner. Changing $\phi\to -u$ and
$\phi^{\dagger}\to-v$  leads to the expression Eq. (\ref{def_S_ot_b}).

\bigskip

\end{document}